\documentclass[preprint2]{aastex6}

\usepackage{color}
\usepackage{graphicx}

\newcommand{\m}{\mathrm}

\begin{document}

\title{Ground-based Transit Observation of the Habitable-zone super-Earth K2-3d}

\author{Akihiko~Fukui\altaffilmark{1}, John~Livingston\altaffilmark{2}, Norio~Narita\altaffilmark{2,3,4}, Teruyuki Hirano\altaffilmark{5},  Masahiro~Onitsuka\altaffilmark{4,6}, Tsuguru~Ryu\altaffilmark{4,6}, and Nobuhiko~Kusakabe\altaffilmark{3,4}}
\email{afukui@oao.nao.ac.jp}

\altaffiltext{1}{Okayama Astrophysical Observatory, National Astronomical Observatory of Japan, NINS, Asakuchi, Okayama 719-0232, Japan}
\altaffiltext{2}{Department of Astronomy, The University of Tokyo, 7-3-1 Hongo, Bunkyo-ku, Tokyo, 113-0033, Japan}
\altaffiltext{3}{Astrobiology Center, NINS, 2-21-1 Osawa, Mitaka, Tokyo, 181-8588, Japan}
\altaffiltext{4}{National Astronomical Observatory of Japan, NINS, 2-21-1 Osawa, Mitaka, Tokyo 181-8588, Japan}
\altaffiltext{5}{Department of Earth and Planetary Sciences, Tokyo Institute of Technology, 2-12-1 Ookayama, Meguro-ku, Tokyo 152-8551, Japan}
\altaffiltext{6}{SOKENDAI (The Graduate University for Advanced Studies), 2-21-1 Osawa, Mitaka, Tokyo 181-8588, Japan}

\begin{abstract}

We report the first ground-based transit observation of K2-3d, a 1.5 $R_\oplus$ planet supposedly within the habitable zone around a bright M-dwarf host star, using the Okayama 188 cm telescope and the multi($grz$)-band imager MuSCAT. Although the depth of the transit (0.7 mmag) is smaller than the photometric precisions (1.2, 0.9, and 1.2 mmag per 60 s for the $g$, $r$, and $z$ bands, respectively), we marginally but consistently identify the transit signal in all three bands, by taking advantage of the transit parameters from K2, and by introducing a novel technique that leverages multi-band information to reduce the systematics caused by second-order extinction. We also revisit previously analyzed {\it Spitzer} transit observations of K2-3d to investigate the possibility of systematic offsets in transit timing, and find that all the timing data can be explained well by a linear ephemeris.
We revise the orbital period of K2-3d to be 44.55612 $\pm$ 0.00021 days, which corrects the predicted transit times for 2019, i.e., the era of the {\it James Webb Space Telescope}, by $\sim$80 minutes.
Our observation demonstrates that (1) even ground-based, 2 m class telescopes can play an important role in refining the transit ephemeris of small-sized, long-period planets, and that (2) a multi-band imager is useful to reduce the systematics of atmospheric origin, in particular for bluer bands and for observations conducted at low-altitude observatories.
 
\end{abstract}


\section{Introduction}

K2, the extended {\it Kepler} mission \citep{2014PASP..126..398H}, is increasing the number of known small transiting planets orbiting nearby M dwarfs, including potentially habitable ones, providing promising targets for future atmospheric studies of small planets by e.g., the {\it James Webb Space Telescope} ({\it JWST}). However, the survey duration of K2 per field ($\sim$80 days) is much shorter than {\it Kepler}'s prime mission (four years), decreasing the number of observable transits for a given period planet by a factor of $\sim$18.
The smaller time baseline per field inhibits precise measurement of the orbital period of transiting planets, as well as transit timing variations (TTVs). Transit followup observations by other telescopes are thus an important way to improve the ephemeris, which is essential in planning future observations, as well as to study the dynamics of multi-planet systems via TTVs.

K2-3d is one such planet worthy of additional followup. K2-3d was discovered around a nearby M0 dwarf in Campaign 1 of the K2 mission \citep[][hereafter C15]{2015ApJ...804...10C}, as the outermost planet among three small planets transiting the same host star. Independent detections of the planetary signal were also reported by \citet{2015MNRAS.454.4159H} and \citet{2016ApJS..222...14V},
and radial velocity measurements of this system were reported by \citet{2015A&A...581L...7A} and \citet{2016ApJ...823..115D}.
K2-3d has an orbital period of 44.6 days, which corresponds to a semi-major axis of 0.208 AU, where the planet receives an incident flux of 1.51$^{+0.57}_{-0.43}$ times that of the Earth (C15). This means that this planet is probably located at the inner edge of \citep[according to the recent Venus model,][]{1993Icar..101..108K} or well within \citep[in the case of synchronous rotation,][]{2014ApJ...787L...2Y,2016ApJ...819...84K} the habitable zone around the M-dwarf host star. 
In addition, the composition of this planet could be dominated by rock, given its radius of 1.52 $^{+0.21}_{-0.20}$ R$_\oplus$ (C15), which is on the boundary between rocky and volatile-rich planets for close-in planets \citep[e.g.,][]{2014ApJ...783L...6W,2015ApJ...801...41R}, although the boundary for cooler planets is less clear.
Given the brightness of the host star ($V=12.17$, $J=9.42$), K2-3d is currently one of the best targets for spectroscopic characterizations of potentially habitable planets.
However, the relatively long orbital period allowed K2 to observe only two transits within its observing campaign, leaving large uncertainties in the transit parameters, particularly in the orbital period.

\citet[][hereafter B16]{2016ApJ...822...39B} observed two additional transits of K2-3d using {\it Spitzer}  (IRAC 4.5~$\mu$m), aiming at improving the transit ephemeris and other parameters. Combining the transit timing data of K2 and {\it Spitzer}, they attempted to reduce the uncertainty of the transit timing prediction in the {\it JWST} era (around 2019 December) from $\sim$6 hr to $\sim$0.4 hr. However, because of the shallow transit depth of K2-3d ($\sim$0.7~mmag), which is similar in amplitude to the systematic noise afflicting {\it Spitzer} light curves, a great deal of caution must be exercised when analyzing the data, and additional transit observations can yield significant further improvement of the ephemeris. Furthermore, there is the potential for dynamic interactions between the planets in the system to cause measurable TTVs, which could only be revealed by additional observations.

In this paper we report the first ground-based transit observation of K2-3d, which was achieved by a 2 m class telescope and a multi-band imager. Although observing the tiny transit of K2-3d is challenging for a ground-based 2 m class telescope, the multi-band simultaneous data help us not only to increase the statistical significance but also to reduce the systematics caused by the Earth's atmospheric extinction.

The rest of this paper is organized as follows. We describe the observation and reduction in Section \ref{sec:obs}. In Section \ref{sec:k2} we reanalyze the K2 data in order to refine the transit parameters, with which we search for the transit signal in the MuSCAT data  in Section \ref{sec:signal_search}. We then reanalyze the {\it Spitzer} data and refine the transit ephemeris in Section \ref{sec:spitzer}. Finally, our results are discussed and summarized in Section \ref{sec:summary}.

\section{Observation and Reduction}
\label{sec:obs}

We conducted a photometric observation of K2-3 on 2016 March 2 UT, when a primary transit of K2-3d was predicted to occur, by using the Multi-color Simultaneous Camera for studying Atmospheres of Transiting exoplanets \citep[MuSCAT,][]{2015JATIS...1d5001N} mounted on the 188~cm telescope at Okayama Astrophysical Observatory (OAO) in Japan. MuSCAT is a three-channel imager for the Sloan $g'_2$, $r'_2$, and $z_\m{s,2}$ bands (hereafter $g$, $r$, and $z$ bands, respectively),
recently developed for validating \citep{2016ApJ...820...41H} and characterizing \citep{2015ApJ...815...47N,2016ApJ...819...27F} transiting planets. 
 MuSCAT provides a field of view of 6\farcm1 $\times$ 6\farcm1 for all channels, within which only two sufficiently bright comparison stars, namely TYC 4923-662-1 ($V=11.88$, $J=10.67$) and GSC2.3 S4RH000119 ($V=13.16$, $J=12.07$), were simultaneously imaged  with the relatively bright target star K2-3.
The exposure time was set to 60~s for all bands. The telescope was defocused such that the FWHM of the stellar point spread function (PSF) was within the ranges of 15--21, 18--24, and 19--25 pixels for the $g$, $r$, and $z$ bands, respectively.  We started the observation at 12:29 UT and continued it through 19:06 UT, which resulted in the total time coverage of 6.6 hr. During the observation the sky was clear, with a waning moon (age of 23 days) for the last two hours.

\begin{figure}
\begin{center}
\includegraphics[width=8cm]{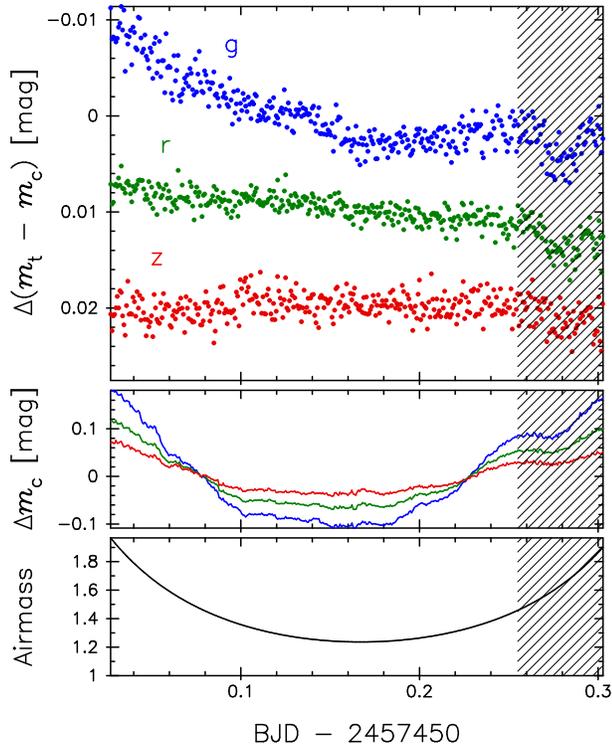}
\caption{
Top: the systematic-uncorrected, differential light curves of K2-3d observed with the 188 cm telescope/MuSCAT. The blue, green, and red light curves are for the $g$, $r$, and $z$ bands, respectively. Each light curve is vertically shifted by an arbitrary amount for clarity. The data in the shaded region are omitted from the analyses due to the uncorrectable systematics. Middle: the median-subtracted raw light curves of the comparison stars. The meanings of the colors are the same as in the top panel. Bottom: theoretical airmass values calculated from the zenith angle of the target star. 
\label{fig:lc_uncorrected}
}
\end{center}
\end{figure}

All the observed images were dark-subtracted and flat-fielded in a standard manner. To create flat-field images, we median-combined 50 dome-flat images for each band that were obtained on the observing night. After the dark-flat correction, we applied a nonlinearlity correction for each CCD. Then aperture photometry was performed for the target and the two comparison stars  by using the customized code \citep{2011PASJ...63..287F}, which calculates the theoretical error for each flux by taking account of the photon noise, detector noise, and scintillation noise.  The aperture radius was optimized for each band such that the dispersion of the differential light curve, which was created by taking the magnitude difference between the target star and the ensemble of the two comparison stars, was minimum. As a result, radii of 20, 22, and 21 pixels were selected for the $g$, $r$, and $z$ bands, respectively. All the time stamps were then converted into barycentric JD (BJD) using the code by \citet{2010PASP..122..935E}. The derived differential light curves are shown in Figure \ref{fig:lc_uncorrected}.
 
In those light curves, we found a 20 minutes  long, 2--4 mmag dip (depending on wavelengths) near the end of the observation (BJD$\sim$2457450.28). We attributed this feature to systematics of the Earth's atmospheric origin rather than an astrophysical phenomenon, because it coincided with a distortion of the raw light curves of both the target and comparison stars (see the middle panel of Figure \ref{fig:lc_uncorrected} for the case of the comparison stars).
This systematics was not correctable, even with the new approach to correct the second-order extinction effect described in Section \ref{sec:new_approach}, and we simply discarded this part (BJD$>$2457450.255, shaded in Figure \ref{fig:lc_uncorrected}) from the rest of the analyses. The numbers of data points to be used for the analyses are 307, 307, and 308 for the $g$, $r$, and $z$ bands, respectively.

\begin{deluxetable}{lcc}
\tablecaption{
Priors for Limb-darkening Parameters.
\label{tbl:limbdark}}
\tablehead{
Instrument & $u_1$ & $u_2$
}
\startdata
K2 & 0.3946 $\pm$ 0.1680 & 0.2607 $\pm$ 0.1170\\
MuSCAT/$g$-band & 0.5779 (fixed) & 0.2017 (fixed)\\
MuSCAT/$r$-band & 0.4877 (fixed)& 0.2592 (fixed)\\
MuSCAT/$z$-band & 0.2529 (fixed)& 0.2734 (fixed)\\
{\it Spitzer} & 0.0477 $\pm$ 0.0332 & 0.1570 $\pm$ 0.0438
\enddata
\end{deluxetable}
\section{Refining the Transit Parameters}
\label{sec:k2}

Because the transit signal is not obvious in the MuSCAT light curves, it is difficult to identify it without any prior information about the shape of the transit light curve. 
We therefore utilize the transit parameters from K2 as the prior information.
Since the original discovery of K2-3d by C15, several new K2 data pipelines have been developed that may achieve higher photometric precision than previous efforts. In our testing we found the light curve for K2-3 produced by the EVEREST pipeline \citep{2016arXiv160700524L} to be of a higher quality than others, as measured by the out-of-transit (OOT) scatter. We therefore update the transit parameters using this light curve.
First, we reproduce the systematic-corrected light curve for K2-3 (EPIC 201367065) using the EVEREST pipeline by modeling the systematics with the transit parts being masked to avoid underestimation of the transit depth, following the suggestion by \citet{2016arXiv160700524L}. Next, we extract individual transit light curves, each covering $\pm$ 5$\times T_{14}$ from the transit center, where $T_{14}$ is the transit duration ($\sim$4 hr). For each light curve we assign rms of the OOT residuals to each flux error bar.
We note that the time-correlated-noise (red-noise) factor $\beta$, which is the ratio of the observed rms for a binned residual light curve to the rms expected from the dispersion of the unbinned residual light curve \citep{2006MNRAS.373..231P,2008ApJ...683.1076W}, becomes unity for both light curves, indicating no sign of red noise.

We then fit the two transit light curves simultaneously using a customized Markov chain Monte Carlo (MCMC) code \citep{2007PASJ...59..763N,2013PASJ...65...27N}. The code uses the transit model by \citet{2009ApJ...690....1O} with the following parameters: the mid-transit time $T_c$ for each transit, the orbital period $P$, the ratio of planetary radius to stellar radius $R_p/R_s$, the scaled semi-major axis $a/R_s$, the transit impact parameter $b$, and the quadratic limb-darkening coefficients $u_1$ and $u_2$.
 We fix $P$ at the value in B16 (44.55765 days) and assume a circular orbit.
For $u_1$ and $u_2$, we let them be free but impose Gaussian priors 
that we determine from the tabulated values of \citet{2012A&A...546A..14C}, with location and scale parameters equal to the mean and twice the standard deviation of all entries that satisfy $3700 \le$ $T_\mathrm{eff} \le 4100 $ K and log$_{10}g \ge 4.5$, respectively. The adopted priors are listed in Table \ref{tbl:limbdark}. Additionally, we impose spectroscopic constraints from C15 in the form of the Gaussian prior $a/R_s \sim 79.6 \pm 10.4$.
Each light curve is also fit with a linear function to take into account the gradient of the baseline:
\begin{eqnarray}
\label{eq:k2}
f(t) = {\cal F_\m{tr}}(t) \times  (k_0 + k_t \Delta t),
\end{eqnarray}
where $f(t)$ is normalized flux at time $t$, $\Delta t \equiv t - T_c'$ is the differential time from a fixed initial guess for the mid-transit time $T_c'$, 
${\cal F_\m{tr}}(t)$ is the transit light-curve model on a flux scale, and $k_0$ and $k_t$ are free parameters. Note that the model flux is computed by integrating time sampling models of 1 minute over the K2 integration time (30 minutes).
We perform five independent MCMC runs with 10$^6$ steps each, and create merged posterior probability distributions for the parameters. The resultant median and 1$\sigma$ uncertainties of $R_p/R_s$, $a/R_s$, $b$, and $T_{14}$ are summarized in Table \ref{tbl:mcmc_k2}. We confirm that our refined values are consistent with those reported by C15 within $\sim$1$\sigma$.

\begin{deluxetable}{lcc}
\tablecaption{
Refined Transit Parameters for K2-3d.
\label{tbl:mcmc_k2}}
\tablehead{
Parameter & This work & C15
}
\startdata
$a/R_s$ & $78.7^{+4.0}_{-5.0}$ & 78.7 $^{+6.7}_{-13.0}$ \\
$b$ & $0.15^{+0.19}_{-0.14}$ & 0.45 $^{+0.23}_{-0.28}$\\
$R_p/R_s$ & $0.02484^{+0.00056}_{-0.00054}$ & 0.0248 $^{+0.0014}_{-0.0010}$\\
$T_{14}$ (hr) & $4.17^{+0.09}_{-0.07}$ & 3.98 $^{+0.17}_{-0.15}$
\enddata
\end{deluxetable}
\section{Extracting the transit signal}
\label{sec:signal_search}

\subsection{1D Search for $T_\m{c}$}
\label{sec:1d_search}

To extract the transit signal from the MuSCAT light curves, we conduct a one-dimensional (1D) search for $T_c$.
In this search we continuously vary $T_c$ with a step size of 0.002 days through the time window [$t_1 - T_{14}/2$, $t_2 + T_{14}/2$], where $t_1$ and $t_2$ are the start and end times of the observation, respectively. At each step we minimize the $\chi^2$ value for a transit+systematic model for each light curve, fixing all the transit parameters except for $T_c$ at either the values derived in Section \ref{sec:k2} ($a/R_s$, $b$, $R_p/R_s$, and $P$) or the theoretical values given by \citet{2012A&A...546A..14C} ($u_1$ and $u_2$, listed in Table \ref{tbl:limbdark}).
For the systematic model, we first apply a simple linear function in the form of
\begin{eqnarray}
\label{eq:standard}
\Delta m_\m{t} = {\cal M_\m{tr}}(t) + k_0 + k_t \Delta t + k_Z Z,
\end{eqnarray}
where $\Delta m_\m{t}$ is the differential magnitude between the target star and the ensemble of the comparison stars, ${\cal M_\m{tr}}$ is the transit light-curve model in magnitude, $Z$ is the theoretical airmass calculated from zenith distance, and $k_0$, $k_t$, and $k_Z$ are coefficients for normalization, time, and airmass, respectively. Note that we do not incorporate any instrumental variables for modeling systematics, such as the stellar displacements on the detectors ($\Delta x$ and $\Delta y$) and FWHM of the stellar PSF, because we find no correlation between the light curves and these variables.

\begin{figure}
\includegraphics[width=8cm]{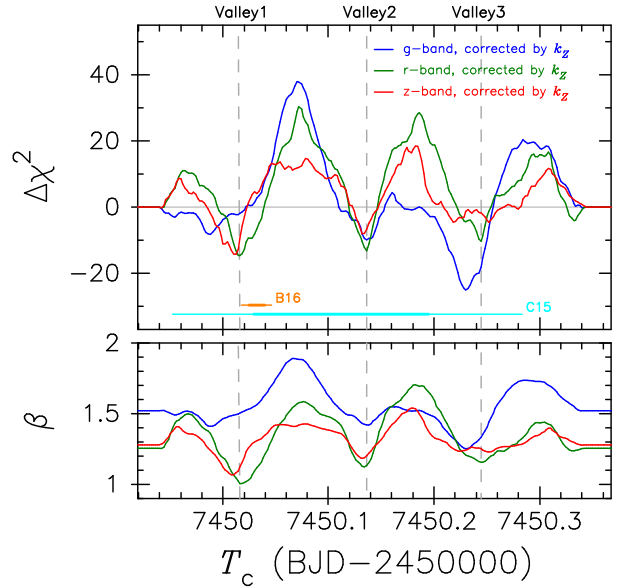}
\caption{
Result of the 1D search for $T_c$. Top: the $\chi^2$ difference between the transit model and the null-transit model as a function of $T_c$. 
The meanings of the blue, green, and red lines are the same as in Figure \ref{fig:lc_uncorrected}.  The thick (thin) orange and cyan horizontal lines represent the 1$\sigma$ (2$\sigma$) range of the expected $T_c$ from B16 and C15, respectively. The gray dashed vertical lines indicate three local minima seen in the $r$-band data, which we label as Valley1 through Valley3 from left to right. Of the three valleys, Valley1 is most likely the true transit signal based on other reductions and lines of evidence. Bottom: the same as the top panel but the $\beta$ value.
\label{fig:pset0}}
\end{figure}

In the top panel of Figure \ref{fig:pset0}, we show the $\chi^2$ difference between the transit and null-transit models, $\Delta \chi^2$, as a function of $T_c$. A negative value means that the transit model is superior to the null-transit one. We note that the absolute (reduced) $\chi^2$ values for the null-transit model are 362.5 (1.19), 346.9 (1.14), and 344.9 (1.13) for the $g$, $r$, and $z$ bands, respectively.
We also calculate the red-noise factor $\beta$, for which we take the median value for the binning sizes ranging from 5 through 20 minutes, as a function of $T_c$. The results are shown in the bottom panel of the same figure.

All three $\Delta \chi^2$ plots show similar patterns, with roughly three local minima (valleys). We mark these valleys (for the $r$ band as a representative) by dashed vertical lines, and label them as Valley1 through Valley3 from left to right. 
Of these, Valley1 gives the minimum $\Delta \chi^2$ values of 14.6 and 14.3 for $r$ and $z$ bands, respectively. The $\beta$ values are also minimal around Valley1 with the values close to unity for the $r$ and $z$ bands. Moreover, the location of Valley1 is marginally ($<$3$\sigma$) consistent with the expected $T_c$ from B16 as indicated by an orange bar in Figure \ref{fig:pset0}. 

\begin{figure*}
\begin{center}
\includegraphics[width=16cm]{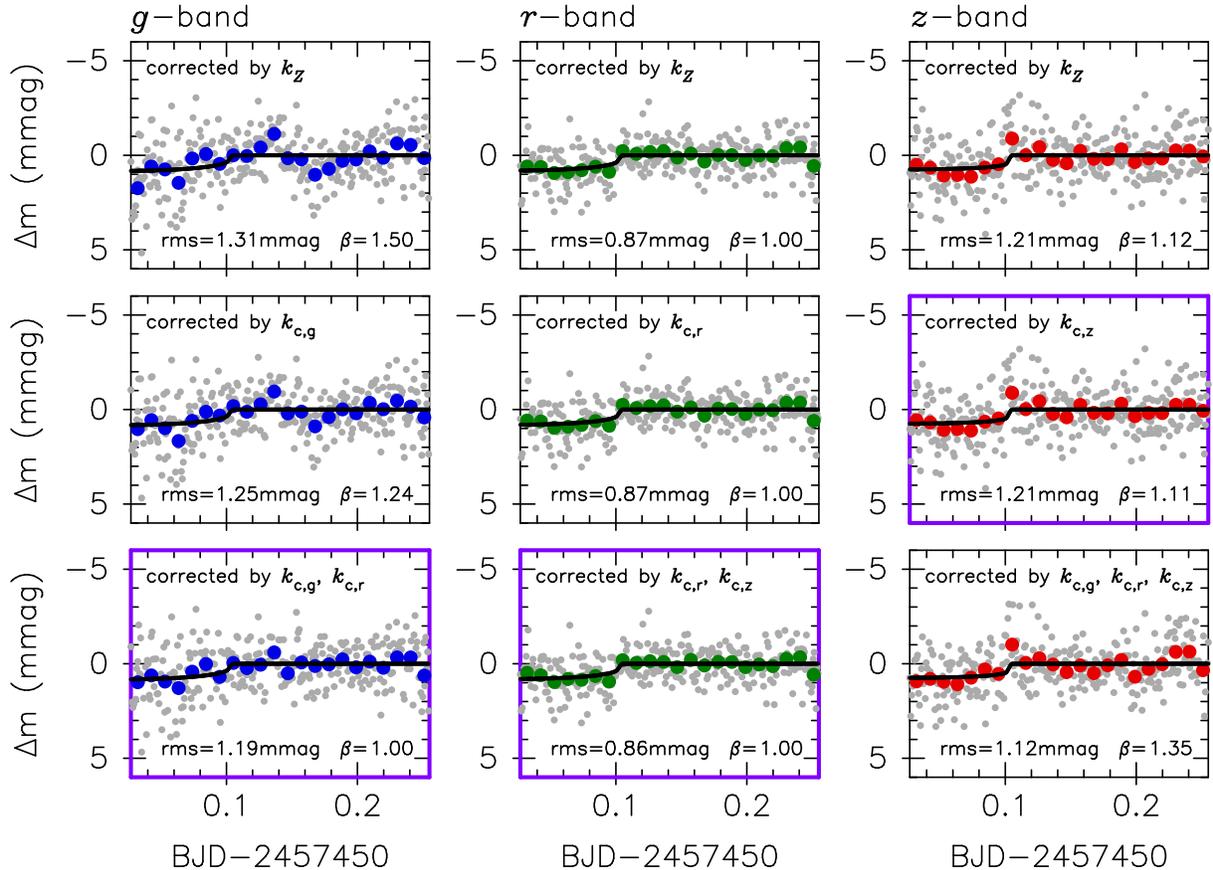}
\caption{
Baseline-corrected light curves of K2-3d. Left, middle, and right columns are for the $g$, $r$, and $z$ bands, respectively. Top, middle, and bottom rows are for the light curves corrected by the conventional method using the theoretical airmass, corrected by Equation (\ref{eq:new1}) using single-band data, and corrected by Equation (\ref{eq:new2}) using multi-band data, respectively.
The small gray and large colored circles represent the unbinned data and the data binned per 15 minutes, respectively. The solid lines represent the transit model at Valley1. The rms and $\beta$ values for each residual light curve are indicated in each panel. The  light curves corrected by the final baseline models are highlighted with a purple frame.
\label{fig:lc_corrected}
}
\end{center}
\end{figure*}

On the other hand, the minimum $\Delta \chi^2$ and $\beta$ values for the $g$ band both appear around Valley3, which is inconsistent with the other bands. However, $\beta$ for the $g$ band is relatively large as a whole ($>$1.2), indicating that the result for this band is less reliable than the other two bands due to the likely higher red noise.
Therefore, at this point Valley1 is the most likely transit signal, although its significance with respect to the other valleys is not high enough, given that the $\chi^2$ difference between Valley1 and Valley2 (Valley3) is 1.4 (4.3) and 6.2 (9.7) for the $r$ and $z$ bands, respectively. 
In the top panels of Figure \ref{fig:lc_corrected} we show the systematic-corrected light curves where we assume that the transit signal is present at Valley1.

\subsection{New Approaches to Correct Second-order Extinction}
\label{sec:new_approach}

The relatively large $\beta$ value for the $g$-band light curve indicates that it is significantly affected by systematic noise. We attribute the majority of the noise to second-order extinction by the Earth's atmosphere. In this section we introduce new approaches to reduce this effect, aiming at enhancing or refuting the evidence for the tentative transit signal. We begin with a brief description of the second-order extinction effect in Section \ref{sec:2nd-order-extinction}, then introduce the new approaches based on single-band data (Section \ref{sec:new1}) and multi-band data (Section \ref{sec:new2}).

\subsubsection{General Description of Second-order Extinction}
\label{sec:2nd-order-extinction}

\begin{figure}
\begin{center}
\includegraphics[width=8cm]{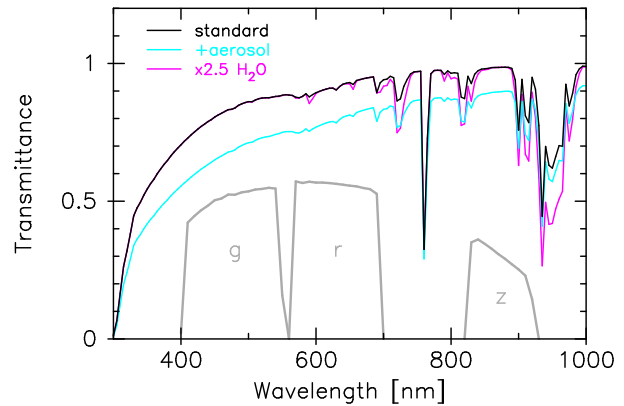}
\caption{
Transmittance of the Earth's atmosphere. The black, cyan, and magenta lines represent a typical atmospheric model for a site at an altitude of 372m in midlatitude without aerosols, that with aerosols, and that with 2.5 times as much H$_2$O. All models are produced by the {\tt libRadtran} code \citep{acp-5-1855-2005}. The gray lines show the total throughputs for MuSCAT including the filter transmittance and the quantum efficiency of the detector. The central wavelength of the atmospheric transmittance for each band can vary with the amounts of aerosols and H$_2$O in the atmosphere.
\label{fig:transmittance}}
\end{center}
\end{figure}

The second-order extinction effect arises when the target and comparison stars have different spectral types.  In such a case, because the extinction coefficient (or the attenuation by the Earth's atmosphere) is a function of wavelength (see Figure \ref{fig:transmittance}), the effective extinction coefficient, which is the weighted average of the extinction coefficient over the passband of a given filter for a given stellar spectrum, differs between the two stars. Consequently, the differential extinction between the two stars varies along with the change in the atmospheric conditions that is dominated by airmass, causing systematic trends in the differential light curves \citep[e.g.,][]{1991PASP..103..221Y,2001PASP..113.1428E,2003MNRAS.339..477B,2011PASP..123.1273M}. 

This effect can be expressed in mathematical form as follows.
The relation between the observed and intrinsic (unattenuated) brightness of the target and comparison stars for a given filter can be written as
\begin{eqnarray}
\label{eq:m_t}
m_\m{t} &=& m_\m{t,0} + A_\m{t} Z + I_\m{t},\\
\label{eq:m_c}
m_\m{c} &=& m_\m{c,0} + A_\m{c} Z + I_\m{c},
\end{eqnarray}
where $m_\m{t}$ ($m_\m{c}$) and $m_\m{t,0}$ ($m_\m{c,0}$) are the observed and intrinsic magnitudes of the target (comparison) star integrated over the passband, respectively, $A_\m{t}$ ($A_\m{c}$) is the effective extinction coefficient for the target (comparison) star at airmass=1, $Z$ is airmass, and $I_\m{t}$ ($I_\m{c}$) is a constant value for the target (comparison) star that is accounted for by the instrumental throughput including the optics and quantum efficiency of the detector.
Assuming that $m_\m{c,0}$ is constant over time, one can express the differential brightness of the target star as
\begin{eqnarray}
\label{eq:delta_m}
\Delta m_\m{t} &\equiv& m_\m{t} - m_\m{c}\nonumber \\
 &=& m_\m{t,0} + (A_\m{t} - A_\m{c}) Z + C,
\end{eqnarray}
where $C$ ($= I_\m{t} - I_\m{c} - m_\m{c,0}$) is a constant. 
The second-order extinction effect arises when $(A_\m{t} - A_\m{c}) Z$ significantly changes during an observation compared with the measurement errors of $\Delta m_\m{t}$.

The effect is expected to be large on our $g$-band observation for the following reasons. 
First, the differential extinction becomes larger at shorter wavelengths where the gradient of atmospheric transmittance is large due to scattering by atmospheric molecules and aerosols (see Figure \ref{fig:transmittance}). Thus, the effect on the $g$ band is generally larger than on redder bands.
Second, the color difference between the target and the ensemble comparison  stars in our observation is moderately large, with $\Delta (V-J) = 1.33$. Such a situation is common for targets that are bright M dwarfs, which are only sparsely distributed on the sky.
 Last, our observation was conducted at a low-altitude (372~m) observatory, where the overhead atmosphere is thicker than at most of the great astronomical sites in the world.
 
Assuming that $A_\m{t}$ and $A_\m{c}$ are constant over the night,
we estimate $A_\m{t}$ and $A_\m{c}$ for each band for our observation by using Equation (\ref{eq:m_t}) and (\ref{eq:m_c}). The results are summarized in Table \ref{tbl:extinction}. We estimate the differential extinction coefficient, ($A_\m{t} - A_\m{c}$), to be -8, 4, and 2 mmag for the $g$, $r$, and $z$ bands, respectively, indicating that the $g$ band gives the largest difference between $A_\m{t}$ and $A_\m{c}$ as expected. Because the airmass value differs by more than 0.7 during the observation, the second-order extinction effect is expected to be considerable in all three bands, which in fact can be seen as long-term trends, prominently in the $g$ band, in Figure \ref{fig:lc_uncorrected}.
We note that although ($A_\m{t} - A_\m{c}$) is expected to be negative for all bands given the color difference between the target and comparison stars, we can attribute the positive values for the $r$ and $z$ bands to the deep TiO absorptions expected to exist around the redder side in each band in the target star's spectrum. 

Such a long-term trend, however, can be corrected at the first order by estimating $(A_\m{t} - A_\m{c})$ using the airmass function, under the assumption that $(A_\m{t} - A_\m{c})$ is constant over the night. This approach is most commonly used and was also applied in Section \ref{sec:1d_search}. 

However, in reality, $A_\m{t}$ and $A_\m{c}$ can temporarily fluctuate due to the temporal changes in atmospheric conditions such as barometric pressure, precipitable water, and the amount and nature of aerosols \citep[e.g.,][]{1994SoPh..152..351H,2007PASP..119.1163S,2010ApJ...720..811B}.
In fact, the raw light curves deviate from a smooth function of airmass (Equation (\ref{eq:m_t}) and (\ref{eq:m_c})) by 25, 16, and 10 mmag in rms for the $g$, $r$, and $z$ bands, respectively (see Table \ref{tbl:extinction}), indicating that the extinction coefficients varied considerably with time with at the level of several per cent.
The fluctuations of $A_\m{t}$ and $A_\m{c}$ could also cause time variations of $(A_\m{t} - A_\m{c})$, which could leave residual systematics that cannot be corrected by the conventional approach using the airmass function.
For example,  fluctuations of 5\% in ($A_\m{t} - A_\m{c}$) could cause systematic signals in the $g$-band light curve with the amplitudes of 0.5--0.7 mmag in the airmass range of $Z$=1.2--1.8, which are comparable to the transit depth of K2-3d ($\sim$0.7 mmag). 

\begin{deluxetable}{ccccccc}
\tablecaption{Effective Extinction Coefficients for Our Observation
\label{tbl:extinction}}
\tablehead{
& \multicolumn{2}{c}{$g$ band} & \multicolumn{2}{c}{$r$ band} & \multicolumn{2}{c}{$z$ band}\\
& Value& rms& Value & rms& Value & rms\\[-4pt]
& (mag) &(mag) & (mag) & (mag) & (mag) & (mag)\\[-12pt]
}
\startdata
$A_\m{t}$ & 0.445 & 0.024 &  0.290 & 0.016 & 0.166 & 0.010\\
$A_\m{c}$ & 0.453 & 0.025 & 0.286 & 0.016 & 0.164 & 0.010\\
($A_\m{t} - A_\m{c}$) &  -0.008 & -- & 0.004 & -- & 0.002 & --\\
\enddata
\end{deluxetable}

\subsubsection{Single-band Approach}
\label{sec:new1}

\begin{figure*}
\begin{center}
\includegraphics[width=16cm]{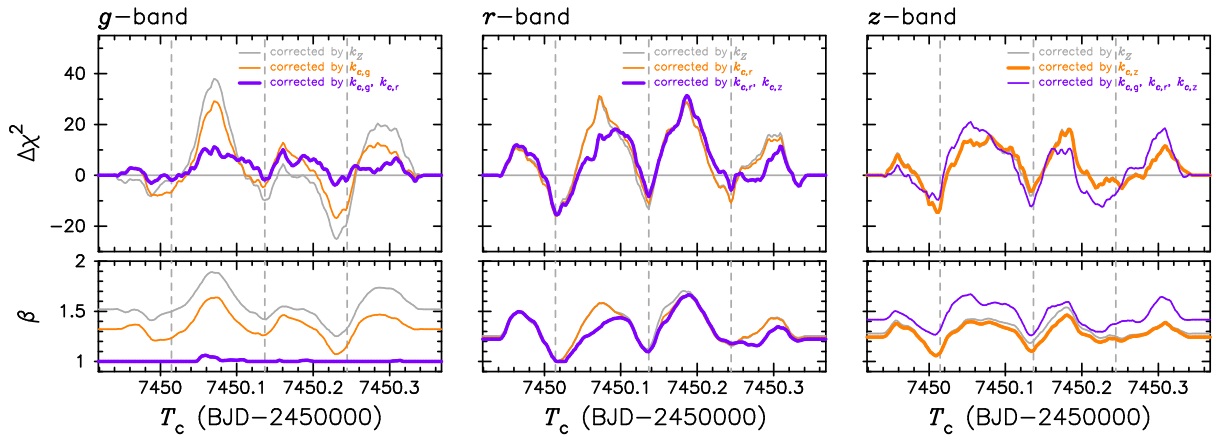}
\caption{
The same as Figure \ref{fig:pset0} but a comparison between the conventional approach and the new approach corrected for second-order extinction. The left, middle, and right panels are for the $g$, $r$, and $z$ bands, respectively. The gray, orange, and purple lines are the results obtained with the conventional (Section \ref{sec:1d_search}), single-band  (Section \ref{sec:new1}), and multi-band (Section \ref{sec:new2}) approaches, respectively. Note that the gray lines are identical to the lines presented in Figure \ref{fig:pset0}. The models selected for further analyses are indicated by bold lines. 
\label{fig:Tc_vs_dchi2_new}}
\end{center}
\end{figure*}

Here we introduce new approaches to correct the systematic noise originating from the second-order extinction. 
First we assume that fluctuations in the atmospheric transparency happen achromatically. In this case the ratio of the effective extinction coefficients for the target and comparison stars is constant over time, such that \begin{eqnarray}
\label{eq:A_t}
A_\m{t} = k_c A_\m{c},
\end{eqnarray}
 where $k_c$ is a constant factor. In this case, Equation (\ref{eq:m_t}) can be transformed into 
\begin{eqnarray}
m_\m{t} = m_\m{t,0} + k_c m_\m{c} + C',
\end{eqnarray}
where $C'$ ($= I_\m{t} - k_c I_\m{c} - k_c m_\m{c,0}$) is a constant.
For a transit light curve, this can be rewritten as
\begin{eqnarray}
\label{eq:new1}
m_\m{t}(t) = {\cal M}_\m{tr}(t) + k_0  + k_t \Delta t + k_c m_\m{c}(t),
\end{eqnarray}
in which we add a linear function of $k_0 + k_t \Delta t$ for the baseline gradient.
This equation differs from Equation (\ref{eq:standard}) in two ways: (1) the raw magnitude of the target star is directly modeled as a function of the raw magnitude of the comparison stars, and (2) it does not rely on the theoretical airmass $Z$.
Using Equation (\ref{eq:new1}), we rerun the 1D search for $T_c$, allowing $k_c$ to be free. For a computational reason, the uncertainty of $m_\m{c}(t)$ in Equation (\ref{eq:new1}) is propagated beforehand to that of $m_\m{t}(t)$ so that we can treat the uncertainty of $m_\m{c}(t)$ as zero during the fitting process.
The results are plotted with orange lines in Figure \ref{fig:Tc_vs_dchi2_new}. 
The effects are visible in the $g$ band; both the depths of the suspicious valleys in the $\Delta \chi^2$ plot and the overall $\beta$ values are slightly reduced.
Nevertheless, the strongest peak is still present around Valley3.
On the other hand, for the $r$ and $z$ bands, we obtain almost identical results to those derived in Section \ref{sec:1d_search}.
In the panels in the middle row of Figure \ref{fig:lc_corrected}, we show the differential light curves that are corrected using the new formula and the transit model that assumes that the transit signal is at Valley1.

\subsubsection{Multi-band Approach}
\label{sec:new2}

Next, we consider a more realistic situation in which the extinction coefficients of $A_\m{t}$ and $A_\m{c}$ are variable in time. 
In general, the variations of the extinction coefficients depend on wavelength.
In other words, the extinction coefficient for the target star for a given band at a given time can in principle be predicted from those for the comparison stars for multiple bands at the same moment. 
We assume here that $A_\m{t}$ for a given band $\lambda$ at time $t$ can be expressed as a linear combination of $A_\m{c}$ for multiple bands at the same time as follows:
\begin{eqnarray}
\label{eq:A_t_lambda}
A_{\m{t},\lambda}(t) = \sum_{\lambda' = \{g, r, z\}} k_{c,\lambda'} A_{\m{c},\lambda'}(t).
\end{eqnarray}
Substituting Equation (\ref{eq:A_t_lambda}) into Equation (\ref{eq:m_t}), one can rewrite Equation (\ref{eq:new1}) as 
\begin{eqnarray}
\label{eq:new2}
m_{\m{t},\lambda}(t) = {\cal M}_\m{tr}(t) &+& k_0 + k_t \Delta t \nonumber \\
&+& \sum_{\lambda' = \{g, r, z\}} k_{c,\lambda'} m_{\m{c},\lambda'} (t).
\end{eqnarray}
Note that this equation is equivalent to Equation (\ref{eq:new1}) when $k_{c,\lambda' (\lambda' \neq \lambda)}=0$. 

The point of this new formula is to correct the extinction for the target star by using the fluxes of the comparison stars in multiple bands. This new approach, however, has to be applied with caution. If the major sources of attenuation between two arbitrary bands are different, then the correlation between the extinction variations for these two bands could be weak. In such a case, incorporating fluxes of different bands in extinction correction could introduce additional, unwanted systematics.
In fact, while scattering by aerosols is the dominant source at shorter wavelengths, water absorption become dominant in redder bands (see Figure \ref{fig:transmittance}), and the  timescale of variability of these two sources could be different.

Keeping this risk in mind, we apply this new formula to the MuSCAT data.
First, to determine which coefficients to include in $k_{c,\lambda}$,
we fit each light curve with different combinations of $k_{c,\lambda}$, assuming that the transit signal is present at Valley1. 
We note that we always include $k_{c,\lambda'}$ when we fit the $\lambda'$-band light curve, because the extinction for the target star for a given band is primarily correlated with that for the comparison stars for the same band. 
In Table \ref{tbl:comp_baseline}, we summarize the statistics of the fittings with different combinations of $k_{c,\lambda}$, in addition to those for the conventional approach (denoted as $k_Z$) for reference. The statistics include Bayesian information criteria BIC $\equiv \chi^2 + k \ln N$, where $k$ is the number of free parameters and $N$ is the number of data points \citep{1978Schwarz}, rms, $\beta$, and rms$\times$$\beta$. 
Although BIC is often used to choose a baseline light-curve model from among several candidates, it does not adequately reflect red noise, which is now of particular concern.
We therefore select the best baseline model such that (1) rms$\times$$\beta$ is minimum, and (2) BIC is minimum if there is another competitive model that has a similar value of rms $\times$ $\beta$ to the minimum one.
The selected combinations are \{$k_{c,g}$, $k_{c,r}$\}, \{$k_{c,r}$, $k_{c,z}$\},  and \{$k_{c,z}$\} for the $g$, $r$, and $z$ bands, respectively. 
We note that even when we conduct the same test assuming that no transit signal is present in the data, we obtain the same combinations of $k_{c,\lambda}$. The statistics for the null-transit model are appended in Table \ref{tbl:comp_baseline}.
This is also true when we assume that a transit signal is present at different valleys.
Therefore, the choice of the baseline models does not produce any bias in the transit detection. 

The fact that the multi-coefficient models are selected for the $g$ and $r$ bands indicates that the multi-band approach is superior to the single-band approach for these two bands. On the other hand, for the $z$ band, the single-coefficient model is selected, meaning that the multi-band approach does not improve but rather degrades the fit. In fact, while the BIC and rms values are reduced by the multi-coefficient models, the $\beta$ value increases substantially so that rms $\times$ $\beta$ increases, implying that in the case of the $z$ band the information from different bands introduces additional systematic noise that outweighs its benefit.

We then rerun the 1D search for $T_c$ with the multi-band approach. Just for comparison, we also try with the coefficients of  \{$k_{c,g}$, $k_{c,r}$, $k_{c,z}$\} for the $z$ band, which is the best combination among the multi-coefficient models.
The results are plotted with purple lines in Figure \ref{fig:Tc_vs_dchi2_new}.
The effect of the multi-band approach is obvious in the $g$ band, where
the suspicious peak around Valley3 is now well suppressed and $\beta$ becomes unity for most $T_c$ values. At the same time, however, no strong peak remains even at Valley1, which is probably a consequence of the fact that the true transit signal is overwhelmed by the photometric noise that is now dominated by statistical noise. Nevertheless, we can still see a possible transit signal in the $g$-band data (see Section \ref{sec:consistency_check}).

 In the case of the $r$ band, the multi-band approach slightly enhances the significance of Valley1 over the other valleys. In the case of the $z$ band, on the other hand, the multi-band approach does not enhance $\Delta \chi^2$ at Valley1 but rather enhances the other valleys, probably due to the additional systematics. 
 In the bottom panels of Figure \ref{fig:lc_corrected}, we show the systematic-corrected light curves from the multi-band approach for the transit model at Valley1.

\begin{table}[t]
\begin{center}
\begin{deluxetable*}{lhcccc|hcccc}
\tablecaption{Comparison of Statistics between the Baseline Models \tablenotemark{a}
\label{tbl:comp_baseline}}

\tablehead{
& \multicolumn{5}{c}{Valley1 ($T_c$ = 2457450.016)} & \multicolumn{5}{c}{Null Transit}\\
\colhead{Model} &  & \colhead{BIC} & \colhead{rms} & \colhead{$\beta$} & \colhead{rms$\times$$\beta$} & & \colhead{BIC} & \colhead{rms} & \colhead{$\beta$} & \colhead{rms$\times$$\beta$} \\ 
\colhead{} & & \colhead{} & \colhead{(mmag)} & \colhead{} &  \colhead{(mmag)} & & \colhead{} & \colhead{(mmag)} & \colhead{} & \colhead{(mmag)} 
} 

\startdata
\sidehead{$g$ band}
$k_{\m{Z}}$ & 360.6 & 377.7 & 1.31 & 1.50 & 1.97 & 362.5 & 379.6 & 1.30 & 1.52  &1.98 \\
$k_{c,g}$ & 330.5 & 347.7 & 1.25 & 1.24 & 1.55 & 336.3 & 353.5 & 1.26 & 1.32 & 1.66 \\
\mbox{\boldmath $k_{c,g} + k_{c,r}$} & {\bf 296.2} & {\bf 319.1} & {\bf 1.19} & {\bf 1.00} & {\bf 1.19} & {\bf 298.0} & {\bf 320.9} & {\bf 1.20} & {\bf 1.00} & {\bf 1.20} \\
$k_{c,g} + k_{c,z}$ & 300.5 & 323.4 & 1.20 & 1.00 & 1.20 & 304.2 & 327.1 & 1.21 & 1.00 & 1.21 \\
$k_{c,g} + k_{c,r} + k_{c,z}$ & 294.9 & 323.5 & 1.19 & 1.00 & 1.19 & 297.0 & 325.7 & 1.20 & 1.00 & 1.20 \\
\sidehead{$r$ band}
$k_Z$ & 332.3 & 349.4 & 0.87 & 1.00 & 0.87 & 346.9 & 364.1 & 0.89 & 1.26 & 1.12 \\
$k_{c,r}$ & 331.3 & 348.5 & 0.87 & 1.00 & 0.87 & 347.3 & 364.5 & 0.89 & 1.24 & 1.10 \\
$k_{c,g} + k_{c,r}$ & 322.2 & 345.1 & 0.85 & 1.21 & 1.03 & 340.2 & 363.1 & 0.88 & 1.46 & 1.28 \\
\mbox{\boldmath $k_{c,r} + k_{c,z}$} & {\bf 325.3} & {\bf 348.2} & {\bf 0.86} & {\bf 1.00} & {\bf 0.86} & {\bf 340.8} & {\bf 363.7} & {\bf 0.88} & {\bf 1.22} & {\bf 1.08} \\
$k_{c,g} + k_{c,r} + k_{c,z}$ & 284.1 & 312.8 & 0.81 & 1.18 & 0.96 & 304.2 & 332.9 & 0.84 & 1.56 & 1.31 \\
\sidehead{$z$ band}
$k_Z$ & 335.4 & 352.6 & 1.21 & 1.12 & 1.36 & 344.9 & 362.1 & 1.23 & 1.28 & 1.58 \\
\mbox{\boldmath $k_{c,z}$} & {\bf 335.8} & {\bf 353.0} & {\bf 1.21} & {\bf 1.11} & {\bf 1.35} & {\bf 345.6} & {\bf 362.8} & {\bf 1.23} & {\bf 1.24} & {\bf 1.53} \\
$k_{c,g} + k_{c,z}$ & 323.2 & 346.1 & 1.19 & 1.35 & 1.60 & 332.1 & 355.0 & 1.21 & 1.46 & 1.76 \\
$k_{c,r} + k_{c,z}$ & 312.9 & 335.8 & 1.17 & 1.44 & 1.68 & 320.8 & 343.7 & 1.18 & 1.55 & 1.83 \\
$k_{c,g} + k_{c,r} + k_{c,z}$ & 287.5 & 316.1 & 1.12 & 1.35 & 1.51 & 292.1 & 320.8 & 1.13 & 1.42 & 1.60 \\
\enddata
\tablenotetext{a}{ The final models are indicated by boldface.}
\end{deluxetable*}
\end{center}
\end{table}

\subsection{$p$-value Calculation}
\label{sec:FAP}

The $\chi^2$ tests in the previous sections indicate that the transit model for Valley1 is nominally preferred over the null-transit model by more than 3$\sigma$ for both the $r$ and $z$ bands. 
However, these calculations do not take into account the effects of systematic noise, and thus probably overestimate the significance of the signal.

To obtain a better estimation of the statistical significance of the signal in the existence of systematics, we calculate the probability that the hypothetical transit signal is observed by random chance under the null hypothesis,
or the $p$-value, for each band as follows.
First, each light curve is fit with a null-transit model, letting only the baseline parameters (the coefficients selected in Section \ref{sec:new_approach}) be free.  Second, the residual light curve is divided into $N$ groups with a bin size of $M$ minutes. Third, the $N$ groups are randomly permuted while the order of the data points in each bin is kept the same so that the patterns of systematic noise are held in each bin. 
Fourth, the permuted residuals are added to the best-fit null-transit model to create a synthetic light curve. Finally, the 1D search for $T_c$ is performed for the synthetic light curve in the same way as in Section \ref{sec:1d_search}. We repeat this procedure 10$^5$ times, randomly changing $M$ in the range from 10 to 20 minutes, which corresponds to the typical time scale of the red noise. We then calculate the probability that the minimum $\Delta \chi^2$ is equal to or less than the observed value for Valley1, namely 1.9, 15.5, and 14.6 for the $g$, $r$, and $z$ bands, respectively.

As a result, we find that the $p$-value is 67.3\%, 3.1 \%, and 3.9\% for the $g$, $r$, and $z$ bands, respectively. These results indicate that, while the $g$-band data have no statistical power, the $r$- and $z$-band data are incompatible with the null hypothesis against the transit hypothesis at significance levels of 2.2$\sigma$ and 2.1$\sigma$, respectively. We also calculate the $p$-value for the ensemble of the $r$- and $z$-band data, in which we observe the minimum $\Delta \chi_\m{sum}^2$ value of 28.2 around Valley1, where $\Delta \chi_\m{sum}^2$ is the sum of $\Delta \chi^2$ for the two bands. In this calculation the order of the permutations is kept the same between the two bands so that the correlations of systematic noise between the two light curves are conserved. The calculated $p$-value is 1.5\%, which corresponds to a significance of 2.4$\sigma$, suggesting that the detection of the transit signal is still marginal.

\subsection{Consistency Check for $T_c$ and $R_p/R_s$}
\label{sec:consistency_check}

\begin{figure}[t]
\includegraphics[width=8cm]{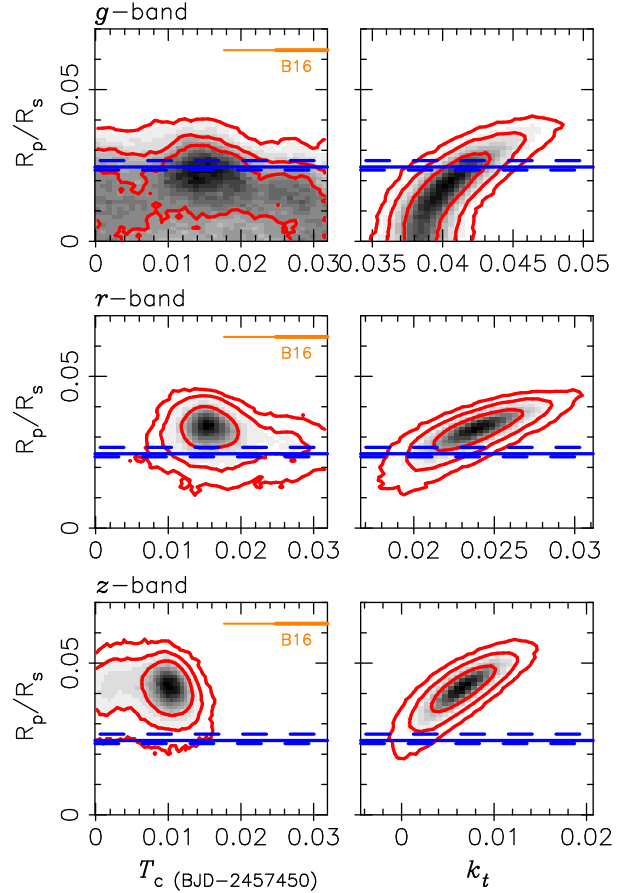}
\caption{Correlation maps between $R_p/R_s$ and $T_c$  (left column), and between $R_p/R_s$ and $k_t$  (right column). The top, middle, and bottom panels are for the $g$-, $r$-, and $z$-band light curves, respectively. The gray maps indicate the density of the posterior probability distribution from the MCMC analyses, and the red contours from inside to outside indicate the 1$\sigma$, 2$\sigma$, and 3$\sigma$ credible regions. The blue solid and dashed lines are the median and 2$\sigma$ upper or lower values, respectively, derived from the K2 data. The thick and thin orange lines represent the 1$\sigma$ and 2$\sigma$ range of the expected $T_c$ from the ephemeris of B16. 
\label{fig:correlation_map}}
\end{figure}

To see the consistency of the tentative transit signal at Valley1 within the three light curves and with what we expect from the past observations, we perform an MCMC analysis for individual light curves.
Before the analysis we rescale the magnitude errors in each light curve such that the reduced $\chi^2$ value for the best-fit transit+systematic model at Valley1 becomes unity and we further rescale them by a factor of $\beta$ (the value for Valley1 listed in Table \ref{tbl:comp_baseline}). During the MCMC analysis we fix $a/R_s$ and $b$ to the values derived from the K2 data in Section \ref{sec:k2}, fix $u_1$ and $u_2$ to the theoretical values, and allow $T_c$, $R_p/R_s$, $k_0$, $k_t$, and the selected coefficients of $k_{c,\lambda}$  to be free. For the $g$ band, to avoid local minima we loosely constrain $T_c$ to the range between 2457450.000 and 2457450.032.
We perform five independent MCMC runs with 10$^6$ steps each for each light curve, and calculate the merged posterior probability distributions of the parameters. The resultant median and 1$\sigma$ uncertainties for $T_c$ and $R_p/R_s$ as well as $k_{c,g}$, $k_{c,r}$, and $k_{c,z}$ are listed in Table \ref{tbl:MCMC_individual}. Figure \ref{fig:correlation_map} shows two-dimensional correlation plots between $R_p/R_s$ and $T_c$ and between $R_p/R_s$ and $k_t$. 
We find that $R_p/R_s$ and $T_c$ for all bands are largely consistent with each other within 2$\sigma$. Remarkably, we detect a marginal signal in the $g$-band data at $\sim$1$\sigma$ level, which has $R_p/R_s$ and $T_c$ values that are consistent with those in the other bands, further supporting the scenario in which the transit signal is present at Valley1. We also find that the $R_p/R_s$ values for the $g$, $r$, and $z$ bands are consistent with the value from K2 (indicated by blue horizontal lines in Figure \ref{fig:correlation_map}) within 1$\sigma$, 2$\sigma$, and 3$\sigma$, respectively. The slightly large discrepancy between the $z$ band and K2 might be caused by systematics in the $z$-band light curve; since only the partial transit was covered by MuSCAT, the transit parameters such as $R_p/R_s$ can easily be affected by systematics. In fact, $R_p/R_s$ is correlated with $k_t$ (the coefficient of the linear function for the baseline) as shown in the right panels in Figure \ref{fig:correlation_map}, indicating that $R_p/R_s$ can be affected by systematics even in the out-of-transit part of the light curves.

Next, to derive a final $T_c$ value and its uncertainty, we fit all the three light curves simultaneously with MCMC.
In order to properly reflect the uncertainties of the {\it a priori} information about the transit shape, we also simultaneously fit the K2 light curves in addition to the MuSCAT light curves.
In this analysis we allow the following parameters to be free: $a/R_s$, $b$, and $R_p/R_s$ as common parameters for all data, $T_c$ for each transit, $u_1$ and $u_2$ for {\it K2 }with the same priors listed in Table \ref{tbl:limbdark}, and the coefficients for the baseline function for each light curve.
For the baseline function we use Equation (\ref{eq:k2}) for the K2 data, and Equation (\ref{eq:new2}) for the MuSCAT data with the selected coefficients in Section \ref{sec:new2}. 
 After performing five independent MCMC runs with 5$\times$10$^6$ steps each, we merge all the MCMC steps and calculate the median and 1$\sigma$ uncertainties, yielding 
\begin{eqnarray}
&&T_c   \m{(BJD_{TDB} - 2450000)}\nonumber \\ 
&&=  7450.0137  \pm 0.0031 \m{(stat.)}  \pm 0.0030 \m{(sys.)},
\end{eqnarray}
where we conservatively add 0.0030 in the uncertainty as a systematic error that corresponds to half of the offset between the best-fit $T_c$ for the individual fit to the $r$- and $z$-band light curves. The estimated $T_c$ is about 25 minutes earlier than predicted from the B16 ephemeris ($T_c = 7450.0319 \pm 0.0071$), although the significance of the discrepancy is only 2.2$\sigma$. We will revisit this issue in Section \ref{sec:spitzer}.

\begin{table}[t]
\begin{deluxetable*}{lccc}
\tablecaption{MCMC Results for the Individual Light Curves.
\label{tbl:MCMC_individual}}
\tablehead{
\colhead{Parameter} & \colhead{$g$ band} & \colhead{$r$ band} & \colhead{$z$ band}
} 
\startdata
$T_c$ & 0.0153 $^{+0.0105}_{-0.0097}$  &  0.0157 $^{+0.0029}_{-0.0024}$ & 0.0096 $^{+0.0020}_{-0.0038}$\\[-2pt]
(BJD-2450000) & & & \\
$R_p/R_s$ & 0.0165 $^{+0.0092}_{-0.0104}$ & 0.0323 $^{+0.0042}_{-0.0048}$ & 0.0417 $^{+0.0049}_{-0.0054}$\\
$k_{c,g}$ & 0.785 $^{+0.030}_{-0.032}$ & 0 (fixed) & 0 (fixed) \\
$k_{c,r}$ & 0.302 $^{+0.050}_{-0.048}$ & 0.961 $\pm$ 0.015 & 0 (fixed)\\
$k_{c,z}$ & 0 (fixed) & 0.060 $\pm$ 0.026 & 0.9925 $\pm$ 0.0047
\enddata
\end{deluxetable*}
\end{table}

\subsection{Short Summary of the Signal Evidence}
Although the statistical significance of the transit signal at Valley1 is marginal, we have gathered the following evidence for it:

\begin{enumerate}
\item The 1D searchs for $T_c$ for the $r$- and $z$-band light curves both give the smallest $\Delta \chi^2$ around Valley1.
\item The $\beta$ values are minimized around Valley1 for both the $r$ and $z$ bands.
\item The $p$-value for the $r$- and $z$-band data against the transit hypothesis is 1.5\%.
\item A consistent signal is also seen in the $g$ band data at 1$\sigma$ level.
\item The measured $T_c$ value is largely consistent with what we expect from B16  within $\sim$2$\sigma$.
\item The $R_p/R_s$ values measured for the $g$, $r$, and $z$ bands are largely consistent with that for K2 within 1$\sigma$, 2$\sigma$, and 3$\sigma$, respectively.
\end{enumerate}

In addition, if the transit signal at Valley1 were a false positive and either the null-transit model or the transit model at the other valleys were true, then the difference in $T_c$ between observed and expected would increase to at least 2 hr, which could not be explained by the gravitational forces from the other two planets in the system (B16). 
Considering all this evidence, the transit signal at Valley1 should be more robust than just the statistical significance estimated as $\Delta \chi^2$ or the $p$-value.

\section{Revisiting the {\it Spitzer} Photometry and Refining the Transit Ephemeris}
\label{sec:spitzer}

Motivated by the discrepancy of 25 minutes between $T_c$ observed by MuSCAT and that expected from B16, we revisit the {\it Spitzer} photometry to investigate the possibility that the orbital period estimated by B16 could have been biased by uncorrected residual systematics in the {\it Spitzer} data. Systematic signals in the {\it Spitzer} photometry are produced by the motion of the PSF on the detector coupled with variations in intrapixel gain.
Because the amplitude of the transit signal from K2-3d is small relative to the systematic signal in the {\it Spitzer} data, imperfect systematic correction can easily cause biases in the transit parameters. 
One clue that this might be the case can be found by comparing the transit duration ($T_{14}$) derived from the K2 light curve with that of {\it Spitzer}; based on the transit parameters reported in B16, $T_{14}$ for {\it Spitzer} is calculated to be 4.67 and 4.44 hours for the transit epochs of $E=6$ and 10, respectively, which are slightly longer than the value for K2 ($4.17^{+0.09}_{-0.07}$ hr).
This might be the result of the conservative approach taken by B16, in which they intended to verify the original discovery by C15 by independently modeling the {\it Spitzer} data using uninformative priors in their MCMC analysis. Furthermore, they modeled the {\it Spitzer} systematics separately from the transit model, so the error bars they report may not accurately reflect the uncertainties in the parameters of the systematics model. 

Because of this, we can hope to improve upon the quality of the extraction of the transit signal from the {\it Spitzer} data by simultaneously fitting the K2 and {\it Spitzer} data with transit+systematic models. 
By simultaneously modeling the K2 data (or alternatively using informative priors), the efficacy of modeling {\it Spitzer}'s systematics may be enhanced due to the additional leverage provided by the constraints on the transit parameters from the K2 data. 

We extract the {\it Spitzer} photometry and model its systematics using the pixel-level decorrelation (PLD) method, following \citet{2015ApJ...805..132D} and B16. PLD uses a linear combination of the individual pixel basis vectors (PBVs) to model the effect of PSF motion on the detector, thus it does not require the calculation of centroids. We use a similar parameterization to B16, with a multiplicative (instead of additive) linear baseline function:
\begin{eqnarray}
\Delta S^t = (\sum\limits_{i=1}^N c_i \hat{P}_i^t + F_\m{tr}(t)) \times (k_0 + k_t \Delta t),
\end{eqnarray}
where the $c_i$ are the PBV coefficients and $\hat{P}_i^t$ is the $i$th pixel value of the normalized pixel grid at time $t$:
\begin{equation}
\hat{P}_i^t = \frac{P_i^t}{\sum\limits_{i=1}^N P_i^t} 
\end{equation}

Instead of using residual rms as a merit function, we select the best circular aperture for the photometry in our analysis by minimizing residual correlated noise, as measured by $\beta$. We find that both the residual rms and residual correlated noise are approximately minimized ($\beta \sim 1$) for a radius of 3 {\it Spitzer}/IRAC pixels in the case of the $E=6$ data. However, for $E=10$, which is more obviously affected by systematics (as seen in the raw light curves), we find that while a similar aperture ($r = 2.9$ pixels) minimizes residual rms, a smaller aperture of radius $r = 2.2$ pixels significantly reduces residual correlated noise ($\beta$ drops from $\sim 1.6$ to $\sim 1.3$). We note that $\beta$ for the $E=10$ data can be further decreased by increasing the complexity of the systematics model, either by using a larger pixel grid (e.g. $5\times5$ vs. $3\times3$) or a higher-order PLD model (e.g. including additional free parameters to allow for second-order terms in the Taylor expansion that PLD is based on). However, our testing of these alternative systematics models showed that the increase in model complexity was not warranted, as determined by BIC. We attempt to compensate for any biases resulting from the time-correlated noise by inflating the photometric uncertainties by their $\beta$ factor. 
Note that rigorous comparisons between the various methods used to correct {\it Spitzer} systematics have been made \citep{2016AJ....152...44I} in which PLD was among the top performers, displaying both high precision and repeatability. However, for any individual data set, one generally high-performing method may be less well suited than another, e.g. if the amplitude of the systematics is unusually high during the observations due to above average thermal instability. A more comprehensive assessment of the {\it Spitzer} K2-3d light curves might entail implementing some of these other methods to check for agreement, but we leave this for a future work.

For the transit model and MCMC we use the open-source Python packages \texttt{PyTransit} \citep{Parviainen2015} and \texttt{emcee} \citep{emcee}. The free parameters of the transit model are $R_p/R_s$, $a/R_s$, $T_c$ for each transit, and the orbital inclination $i$.
We assume a circular orbit and a quadratic limb-darkening law. We impose Gaussian priors on the limb-darkening coefficients, which are determined in the same manner as in Section \ref{sec:k2} and are listed in Table \ref{tbl:limbdark}. We also impose the Gaussian prior $a/R_s \sim 79.5 \pm 10.4$ in the same way as in Section \ref{sec:k2}.
Note that we ignore any wavelength dependence of $R_p/R_s$ caused by a possible atmosphere, which is beyond the sensitivity of the data we model in this work.

\begin{figure}[t]
\vspace{20pt}
\includegraphics[width=8cm]{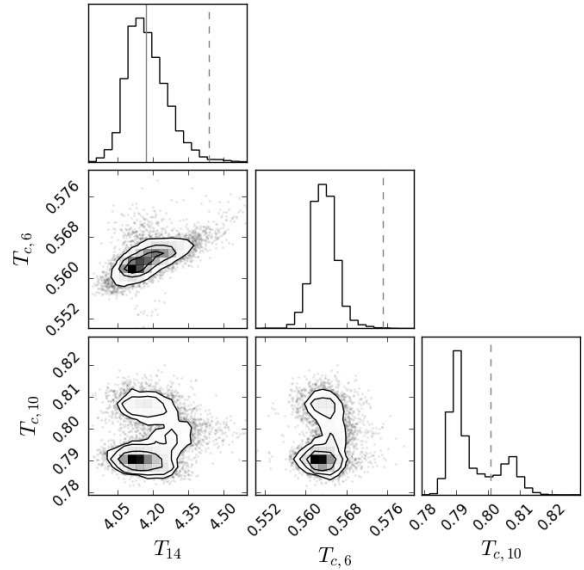}
\caption{Corner plot showing the PPDs of $T_{14}$ [hours], $T_{c,6}-2457093$ [BJD], and $T_{c,10}-2457271$ [BJD] (along the diagonal), and the correlations between them (off-diagonal). The solid line indicates the K2-only value we find for $T_{14}$, and the dashed lines indicate the values for $T_{14}$ ($E=10$), $T_{c,6}$, and $T_{c,10}$ reported by B16.
\label{fig:T14_Tc}
}
\end{figure}

In Figure \ref{fig:T14_Tc} we present a corner plot showing the posterior probability distributions (PPDs) of $T_{14}$, $T_{c,6}$ and $T_{c,10}$, where $T_{c,6}$ and $T_{c,10}$ are $T_c$ for $E=6$ and 10, respectively, and the correlations between them. $T_{14}$ converged to $4.166^{+0.096}_{-0.070}$ hr, which is fully consistent with that from the K2-only fit but is significantly shorter than the values from B16. This indicates that the transit fits by B16 were most likely biased by systematics that we are now able to correct by simultaneously fitting the K2 data and systematics model. 

\begin{figure}[t]
\vspace{20pt}
\includegraphics[width=8cm]{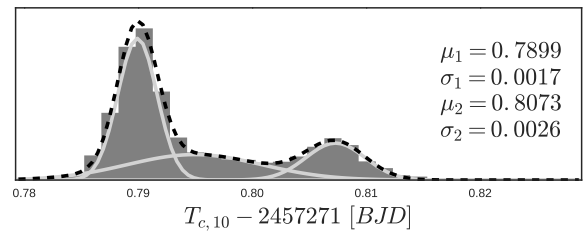}
\caption{The posterior probability distribution of $T_{c,10}$ fit with a triple Gaussian, each component of which is indicated with a gray curve. The dashed line indicates the sum of the individual components, and the parameters of the two modes with the highest probability are displayed.
\label{fig:PPD_Tc}
}
\end{figure}

$T_{c,6}$ also converges to a unimodal distribution with a median value and 1$\sigma$ credible interval of  $T_{c,6} = 7093.5634 ^{+0.0024}_{-0.0020}$, which is earlier than the value of B16 by $\sim$2$\sigma$. This discrepancy could arise from the same cause as the shrinkage of $T_{14}$, as indicated by the slight positive correlation between $T_{14}$ and $T_{c,6}$ (see Figure \ref{fig:T14_Tc}). On the other hand, the PPD for $T_{c,10}$ is significantly bimodal ($T_c-2457271\sim$ 0.790 and 0.807), with more probability mass in the earlier mode. This can be interpreted as the result of the higher levels of ``uncorrectable'' systematics in the data set, as evident  from the $\beta$ factor. We find that the $T_{c,10}$ PPD can be well fit by a triple Gaussian model with two higher-amplitude modes and a low-amplitude mode in between, as shown in Figure \ref{fig:PPD_Tc}. This analysis yields a mean and standard deviation of ($\mu_1$, $\sigma_1$) = (0.7899, 0.0017) and ($\mu_2$, $\sigma_2$) = (0.8073, 0.0026) for the dominant modes. In table \ref{tbl:TTs} we summarize the $T_{c}$ values for the available five transits estimated in this work, along with the values reported in the literature. 

\begin{figure}[t]
\begin{center}
\includegraphics[width=8cm]{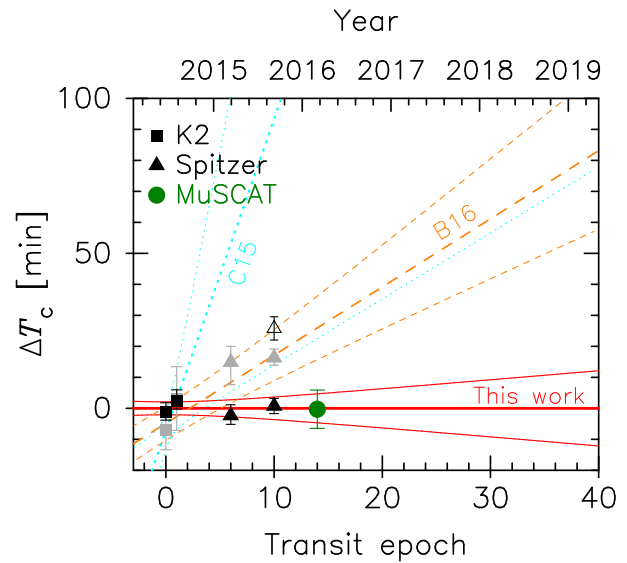}
\caption{
Residuals of the observed $T_c$ from the refined linear ephemeris (red bold line). The red thin lines represent 1$\sigma$ uncertainties of the ephemeris. The black squares, black triangles, and green circle are for K2,  {\it Spitzer}, and MuSCAT, respectively, refined or newly measured in this work. The open triangle represents the second solution for $T_{c,10}$, which is most likely a false positive signal. The gray squares and triangles represent the values from C15 and B16, respectively. The cyan dotted and orange dashed lines show the transit ephemerides (bold) and their 1$\sigma$ uncertainties (thin) provided by C15 and B16, respectively.
\label{fig:TTs}}
\end{center}
\end{figure}

When we choose the mean value of the highest amplitude (earlier) mode of the $T_{c,10}$ PPD, all five $T_c$ from this work are well aligned, giving a $\chi^2$ value for the best-fit linear function of 1.41 over three degrees of freedom. We obtain the refined linear transit ephemeris of
\begin{eqnarray}
T_c  = 6826.2282\ (14)  + 44.55612\ (21) \times E,
\end{eqnarray}
 where the number in parentheses represents the last two digits of 1$\sigma$ uncertainty. 
The residuals of $T_c$ from the best-fit linear function along with the ephemerides of previous work are shown in Figure \ref{fig:TTs}. 

On the other hand, when we choose the mean value of the second-highest amplitude (later) mode of the $T_{c,10}$ PPD, $\chi^2$ becomes 30.1, with a maximum dispersion of $\sim$15 minutes from the linear ephemeris. Although these dispersions might be caused by the gravitational forces from other planets, it is unlikely, considering the orbital architecture of this system (at least to our current knowledge), for K2-3d to show TTVs with an amplitude of more than 5 minutes (B16). We therefore argue that the earlier $T_{c,10}$ with a linear ephemeris is the preferred scenario under Occam's razor, i.e., the simplest explanation should be selected among competing hypotheses. Furthermore, analysis of the $E=10$ light curve resulting from an aperture of larger radius ($r = 2.9$ pixels), which minimizes residual rms instead of correlated noise, produced a bimodal PPD for $T_{c,10}$ with higher probability mass in the later mode; we interpret this as evidence that the later mode is positively correlated with the strength of uncorrected residual systematics, and is therefore less likely to be the result of the transit signal.

\begin{deluxetable*}{llcccc}
\tablecaption{List of observed $T_c$ for K2-3d.
\label{tbl:TTs}
}
\tablehead{
Epoch & Instrument & \multicolumn{3}{c}{$T_c$(BJD)-2450000}\\
& & This Work & C15 & B16
}
\startdata
0 & K2 & $6826.2274 \pm 0.0020$ & $6826.2232 ^{+0.0037}_{-0.0043}$ & -- \\
1 & K2 & $6870.7860 ^{+0.0024}_{-0.0019}$ & $6870.7863 ^{+0.0073}_{-0.0070}$ & -- \\
6 & {\it Spitzer} & $7093.5633 ^{+0.0024}_{-0.0020}$ & -- & $7093.5752 ^{+0.0036}_{-0.0050}$\\
10 & {\it Spitzer} & $7271.7899 \pm 0.0017$ (best solution) & -- & $7271.8007 ^{+0.0019}_{-0.0017}$\\
& & $7271.8073 \pm 0.0026$ (2nd solution) & &\\
14 & MuSCAT & $7450.0137 \pm 0.0043$ $\tablenotemark{a}$ & -- & --
\enddata
\tablenotetext{a}{ Both the statistical and systematic errors are added quadratically.} 
\end{deluxetable*}

\section{Discussion and Summary}
\label{sec:summary}

We have conducted the first ground-based transit observation of K2-3d, a planet of 1.5 Earth radii that is considered to be inside the habitable zone, using the OAO 188 cm telescope and the multi-band imager MuSCAT. Although the depth of the transit is only 0.7 mmag, which is smaller than the photometric precision per exposure and vulnerable to systematics, we have marginally but consistently detected the transit signal thanks to a priori knowledge of the transit shape from the K2 light curves and the multi-band observations with the new extinction-correction approaches.

We have also revisited the {\it Spitzer} data obtained by B16.
Unlike B16, we have simultaneously fit both the {\it Spitzer} and K2 data with both the  transit model and {\it Spitzer}'s systematic models, which has allowed us to reduce the residual systematics that biased the results of B16. Although the posterior distribution of $T_{c,10}$ becomes bimodal, the earlier solution has a higher likelihood and can explain all the transit timing data including the MuSCAT one well by a linear ephemeris.
Adopting the earlier solution for $T_{c,10}$, we have refined the orbital period to be $44.55612 \pm 0.00021$ days, which is 130 s shorter than the value in B16. This discrepancy would accumulate to $\sim$80 minutes in the predicted transit times in 2019, and is thus critical for planning future followup observations by, e.g., {\it JWST}. This work highlights the importance of a careful analysis for a tiny transit signal that is comparable to the systematic signals, as well as the importance of independent followup observations to verify the previous results.

The transit depth of K2-3d is the second shallowest among those observed by ground-based telescopes, following 55~Cnc~e \citep[0.4 mmag transit observed by the 2.5m Nordic Optical Telescope,][]{2014ApJ...797L..21D}.
Our observation demonstrates that ground-based photometric observations can play an important role in improving the transit ephemeris of small-sized, long-period planets, including potentially habitable ones, discovered by on-going and future space-based transit surveys such as K2 and the {\it Transiting Exoplanet Survey Satellite} \citep{2015JATIS...1a4003R}, whose survey durations are limited.

We have also demonstrated that a multi-band imager like MuSCAT is useful not only for obtaining multi-band light curves simultaneously but also for reducing the systematic noise that originates from the second-order extinction effect, especially for shorter wavelengths and for observations conducted at low-altitude observatories. 
This means that our new technique will also be useful in studying planetary atmospheres via transmission spectrophotometry, because the shorter wavelengths in the optical region are important for seeing the effect of Rayleigh scattering \citep[e.g.,][]{2008MNRAS.385..109P}. 
Note, however, that there can be multiple timescale and wavelength dependencies in the variations of the differential extinction coefficient ($A_\m{t} - A_\m{c}$), because there are various sources of variations in extinction, including the temporal changes of pressure, winds, the nature of aerosols, and precipitable water. In fact, an uncorrectable dip was seen near the end of the MuSCAT light curves (see Figure \ref{fig:lc_uncorrected}), which might be caused by variations of ($A_\m{t} - A_\m{c}$) on a different timescale. These complex systematic features cannot be corrected by the simple form of Equation (\ref{eq:new2}), and require higher-order models or non-parametric approaches such as Gaussian processes \citep[e.g.,][]{Rasmussen_A_2006,2012MNRAS.419.2683G}, which is left as future work.

\acknowledgements
This work was supported by the Astrobiology Center Project of  National Institutes of Natural Sciences (NINS) (Grant Numbers AB281012 and JY280092).
This work was also supported by JSPS KAKENHI (Grant Numbers JP25247026 and JP16K17660).

\bibliographystyle{aasjournal}
\bibliography{ref}

\end{document}